\documentclass[final]{aipproc}
\layoutstyle{6x9}

\begin{document}                                              

\title{Near-barrier Fusion Induced by Stable Weakly Bound and 
Exotic Halo Light Nuclei}



\author{C.~Beck}{
   address={Institut Pluridisciplinaire Hubert Curien, UMR7178, CNRS-IN2P3 et
Universit\'{e} Louis Pasteur (Strasbourg I), B.P. 28,
F-67037 Strasbourg Cedex 2, France}
}

\author{A.~S\`anchez i Zafra}{
   address={Institut Pluridisciplinaire Hubert Curien, UMR7178, CNRS-IN2P3 et
Universit\'{e} Louis Pasteur (Strasbourg I), B.P. 28,
F-67037 Strasbourg Cedex 2, France}
}

\author{A. Diaz-Torres}{
   address={The Australian National University, Canberra ACT 0200,
Australia}
}

\author{I.J.~Thompson}{
   address={Department of Physics, University of Surrey, Guildford GU2 7XH,
U.K.}
}

\author{N.~Keeley}{
   address={DSM/DAPNIA/SPhN CEA Saclay, F-91190 Gif-sur-Yvette, France}
}

\author{F.A.~Souza}{
   address={Departamento de Fisica Nuclear, Laboratorio Pelletron, S\~ao
Paulo, Brazil}
}

\begin{abstract}
The effect of breakup is investigated for the medium 
weight $^{6}$Li+$^{59}$Co system in the vicinity of the Coulomb barrier.
The strong coupling of breakup/transfer channels to fusion is
discussed within a comparison of predictions of the Continuum Discretized
Coupled-Channels model which is also applied to $^{6}$He+$^{59}$Co a
reaction induced by the borromean halo nucleus $^{6}$He.
\end{abstract}

\maketitle

\section{Introduction}

In reactions with weakly bound nuclei, the influence on the fusion process of
coupling both to collective degrees of freedom and to breakup/transfer
channels is a key point for the understanding of N-body systems in quantum
dynamics. The diffuse cloud of neutrons of halo nuclei was expected to lead
to significant enhancement of the fusion cross section at sub-barrier energies 
as compared to predictions of one-dimensional barrier penetration
models~\cite{Canto06}. This was understood in terms of the dynamical processes
arising from strong couplings to collective inelastic excitations of the target
and projectile~\cite{Canto06,Liang06}. However, in the case of reactions where
at least one of the colliding nuclei has a sufficiently low binding energy for
breakup to become a competitive process, conflicting model predictions and
experimental results were reported~\cite{Canto06,Liang06}. Recent experimental
results with $^{6,8}$He beams show that the halo of $^{6}$He does not enhance
the fusion probability, confirming the prominent role of neutron transfer in
$^{6}$He induced fusion reactions~\cite{Canto06,Raabe04,DeYoung05,Penion06}. 
The effect of non-conventional transfer/stripping processes appears to be less
significant for stable weakly bound projectiles~\cite{Beck03,Beck04} on
medium-mass target as compared to $^{208}$Pb~\cite{Signorini03}. 

Excitation functions for sub- and near-barrier total (complete + incomplete) 
fusion cross sections measured for the $^{6,7}$Li+$^{59}$Co 
reactions~\cite{Beck03} were compared to Continuum-Discretized 
Coupled-Channels ({\sc CDCC}) calculations~\cite{Diaz02} indicating only a 
small enhancement of total fusion for the more weakly bound $^{6}$Li 
below the Coulomb barrier, with similar cross sections for both reactions 
at and above the barrier~\cite{Diaz03}. This result is consistent with 
the rather low breakup cross sections measured for the $^{6}$Li+$^{59}$Co 
reaction even at incident energies larger than the Coulomb 
barrier~\cite{Souza04}.

In this work we present {\sc CDCC} calculations for elastic scattering
(including extracted total reaction cross sections), total fusion, and breakup
of weakly bound stable ($^{6}$Li and $^{7}$Li) and radioactive ($^{6}$He) light
projectiles from a medium-mass target ($^{59}$Co). As far as exotic halo
projectiles are concerned, a systematic study of $^{4,6}$He induced fusion
reactions~\cite{Beck04} with an improved three-body {\sc CDCC}
method~\cite{Diaz02,Diaz03} using a dineutron model for $^{6}$He
($\alpha$-$^2$n) is in progress. Some preliminary results on total fusion
of $^{4}$He and $^{6}$He with $^{59}$Co will be presented for the first time in
the last Section of the paper. 

\section{{\sc CDCC} description of $^{6,7}$Li+$^{59}$Co elastic scatterings}

In the present work, detailed {\sc CDCC} calculations for the interaction 
of $^{6,7}$Li on the medium-mass target $^{59}$Co are applied in order to 
provide a simultaneous description of elastic scattering, fusion as 
well as breakup.

Details of the calculations concerning the breakup space (number of 
partial waves, resonances energies and widths, maximum continuum energy 
cutoff, potentials, ...) have been given in previous publications
\cite{Diaz02,Diaz03} (in particular in Tables I, II and III of 
\cite{Diaz03}). The {\sc CDCC} scheme is available in the general coupled 
channels (CC) code {\sc FRESCO} \cite{Fresco}. 

Before investigating whether the proposed {\sc CDCC} formalism can be 
also applied to halo nuclei such as $^{6}$He, we present the full 
description of the $^{6}$Li $\rightarrow$ $\alpha$+$d$ and $^{7}$Li 
$\rightarrow$ $\alpha$+$t$ clusters as two-body objects, respectively. 
In the fusion calculations the imaginary parts of the off-diagonal 
couplings were neglected, while the diagonal couplings included
imaginary parts~\cite{Diaz03}. Otherwise full continuum couplings have 
been taken into account so as to reproduce the elastic scattering 
data~\cite{Beck04,Souza04}. We have used short-range imaginary fusion 
potentials for each fragment separately. This is equivalent to the use 
of incoming wave boundary conditions in {\sc Ccfull} calculations
\cite{Beck03}.

      \begin{figure}
        \includegraphics[height=12cm]{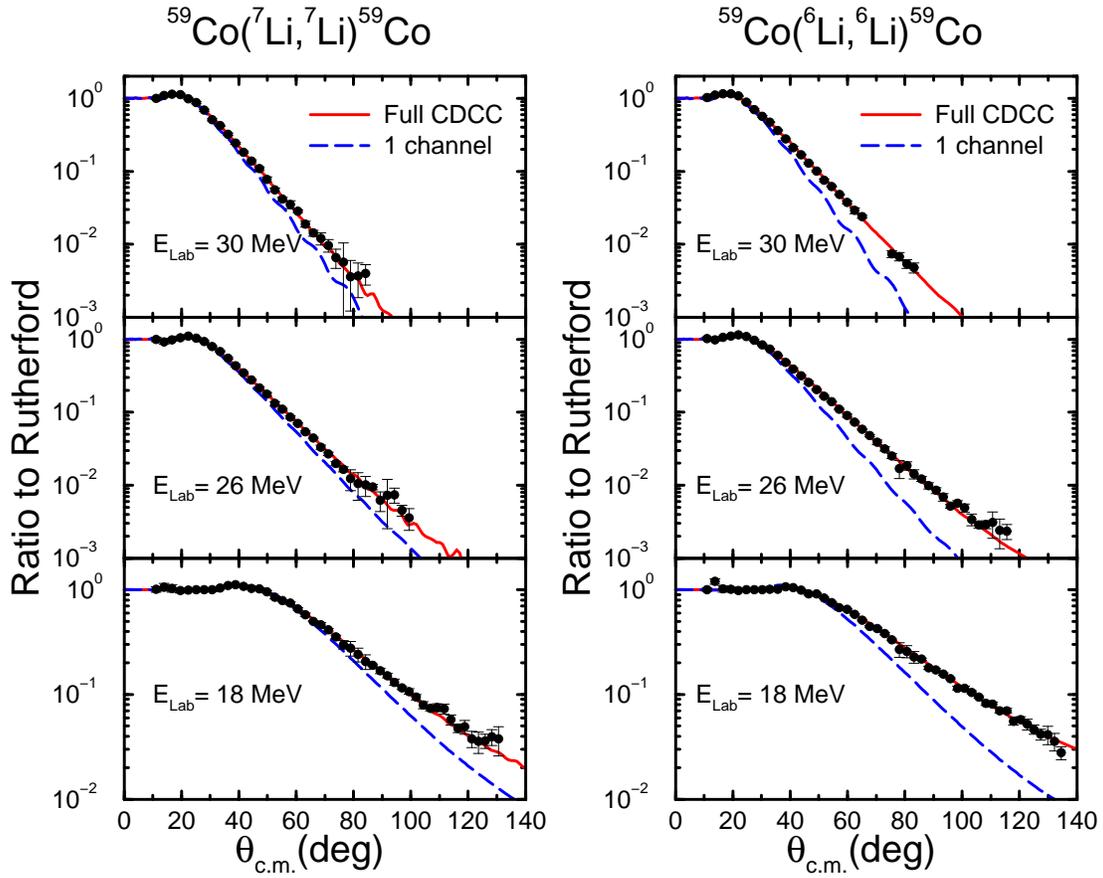}
        {\caption{\label{}{\small Elastic scattering for
                              $^{7}$Li+$^{59}$Co~\cite{Beck04} (left panel)
                              and $^{6}$Li+$^{59}$Co~\cite{Beck04} (right
                              panel). The curves correspond to {\sc CDCC}
                              calculations with (solid lines) or without
                              (dashed lines) couplings with the continuum
                              as discussed in the text.}}}
       \end{figure}

Results of the comparison of the {\sc CDCC} calculations for the elastic
scattering with data of Ref.~\cite{Beck04,Souza04} are shown in Fig.~1 
for $^{7}$Li+$^{59}$Co (left panel) and $^{6}$Li+$^{59}$Co (right panel), 
respectively. The two different curves are the results of calculations 
performed with (solid lines) and without (dashed lines) 
$^{6,7}$Li $\rightarrow$ $\alpha$ + $d,t$ breakup couplings with the 
continuum (i.e. continuum couplings).

       \begin{figure}
       \includegraphics[height=12cm]{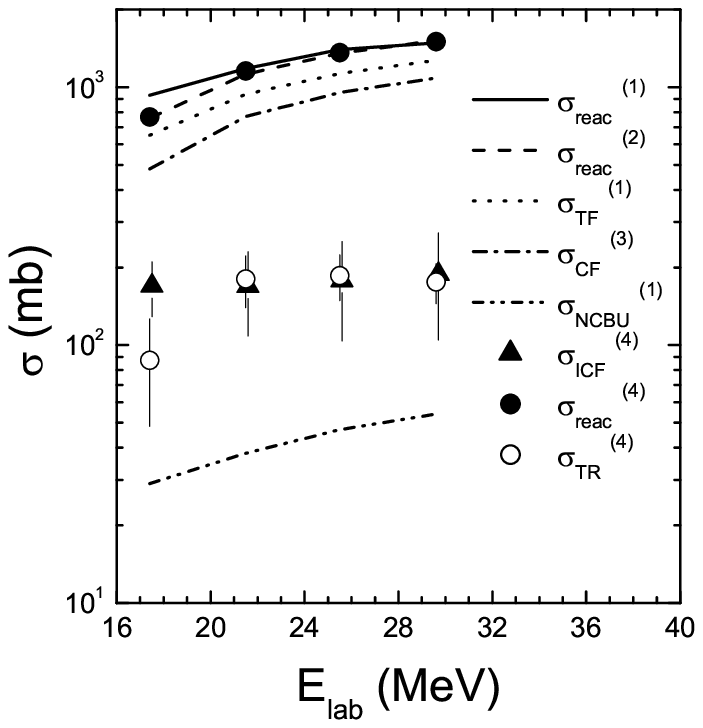}
      {\caption{\label{}{\small $^{6}$Li+$^{59}$Co      
                             excitation functions evaluated with (1)
                              {\sc CDCC} (2) OM fits~\cite{Souza04}
                              (3) CF extrapolations and (4) TR data
                              \cite{Souza06} and ICF = TF-CF~\cite{Souza06}.}}}
        \end{figure}

It is interesting to note that the initial optical-model analysis (OM)
adopted by Souza and collaborators~\cite{Souza04} was found to be ambiguous
for the two lowest energies when using a parameter-free nonlocal potential.
The agreement between the full calculations and data is very good. A 
similar comparison has been provided for the elastic scattering of both 
$^{7}$Li+$^{65}$Cu and $^{6}$Li+$^{65}$Cu reactions~\cite{Shrivastava06}. 
The effect of breakup on elastic scattering, stronger for $^{6}$Li as 
expected, is illustrated by the difference between the one-channel 
calculations (comparable to the OM calculations~\cite{Souza04}) and 
the full {\sc CDCC} results. 

The same {\sc CDCC} description that uses potentials (similar to OM Potentials
of \cite{Souza04}) to 
fit the measured elastic scattering angular distributions~\cite{Beck04,Souza04} 
of Fig.~1 permits one to calculate total reaction cross sections (full curve) 
and non capture breakup (NCBU) yields, as defined in \cite{Canto06}) and 
plotted in Fig.~2 (curve labelled NCBU) with a comparison with the
$^{6}$Li+$^{59}$Co data of Ref.~\cite{Souza06}. The effect of the 
$^{6}$Li breakup and its competition with other reaction mechanisms is 
discussed more deeply in the following Section.

\section{Full {\sc CDCC} description of breakup for $^{6}$Li+$^{59}$Co}

The total calculated breakup cross sections, plotted in Fig.~2 by the 
dashed-double-dotted curve labelled NCBU~\cite{Souza06}, were obtained by 
integrating contributions from the states in the continuum up to 8 MeV.
They are found to be rather small compared with total fusion (TF) {\sc CDCC}
cross sections ~\cite{Diaz03} (dotted line) or with complete fusion (CF) 
cross sections (dotted-dashed line) extrapolated from published TF 
data~\cite{Beck03}. These large discrepancies have also been observed
for the $^{6}$Li+$^{208}$Pb reaction~\cite{Signorini03}.

The total reaction cross sections obtained from fits with OM 
potentials~\cite{Souza04} (dashed line with black points for $^{6}$Li in 
Fig.~2 but not shown for $^{7}$Li) and {\sc CDCC} calculations (solid line) 
are mostly dominated by TF cross sections. Their cross section ratios 
confirm the observed small enhancement of TF cross section for the more 
weakly bound $^{6}$Li nucleus at sub-barrier energies~\cite{Beck03}.  
Similar yields were measured for both reactions at and above the Coulomb 
barrier~\cite{Beck03} in concordance with {\sc CDCC} calculations
\cite{Diaz03} for both TF and total reaction cross sections.

Although the calculated values are significantly smaller than CF and other
measured cross sections (incomplete fusion ICF and transfer TR) for all 
energies, the previous analysis of the $^{6}$Li+$^{59}$Co reaction appears 
to be quite comprehensive when most of the cross sections are compared 
in a consistent way. However, it is still to be determined how much of TR 
yields are included in the so-called measured breakup cross section.

       \begin{figure}
       \includegraphics[height=10cm]{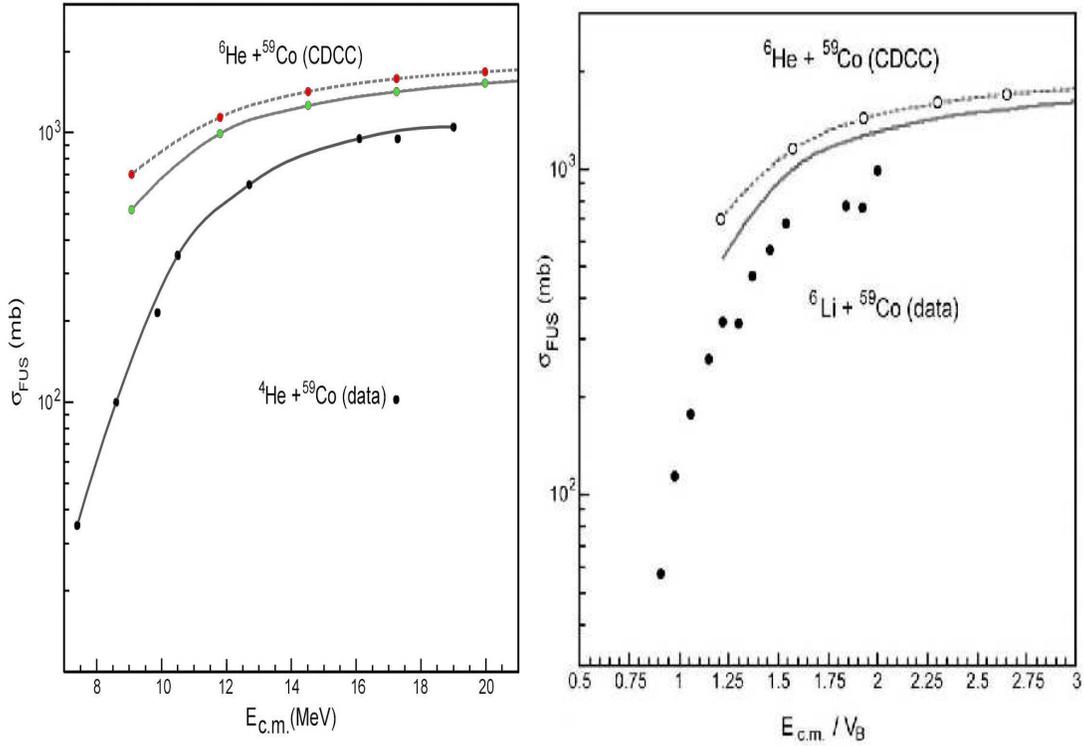}
                      {\caption{\label{}{\small Total fusion excitation
                              functions for $^{4}$He+$^{59}$Co (data points
                              \cite{McMahan80} and solid black line for
                              {\sc CC} predictions on left panel) for 
                              $^{6}$Li+$^{59}$Co (data points~\cite{Beck03}
                              on right panel), and for
                              $^{6}$He+$^{59}$Co. The curves correspond to
                              {\sc CDCC} calculations for  $^{6}$He+$^{59}$Co
                              with (dashed line) or without (thin
                              line) couplings to the continuum.}}}
        \end{figure}

\section{{\sc CDCC} description of $^{6}$He+$^{59}$Co}

In the following we present similar calculations applied to the 
two-neutron halo nucleus $^{6}$He. The present case is much more 
complicated since $^{6}$He breaks into three fragments ($\alpha$+n+n) 
instead of two ($\alpha$+d), and the {\sc CDCC} method is in current 
development for two-nucleon halo nuclei \cite{Matsumoto04,Thompson05}. 
Hence a dineutron model~\cite{Rusek05} is adopted for the 
$^{6}$He+$^{59}$Co reaction: i.e. we assume a two-body cluster 
structure of $^{6}$He = $^{4}$He+$^{2}$n with an $\alpha$-particle core 
coupled to a single particle, a dineutron ($^{2}$n). Couplings to 
resonant (2$^{+}$, E$_{ex}$ = 0.826 MeV) and non-resonant continuum 
states (up to f-waves) are included. The fact that the dineutron is not 
an object with both fixed size and fixed energy (Heisenberg principle) 
might be a critical point in the present model. 

Results of the {\sc CDCC} calculations for TF of the $^{6}$He+$^{59}$Co system
compared to $^{4}$He+$^{59}$Co and $^{6}$Li+$^{59}$Co are shown in Fig.~3.
On the left panel of Fig.~3 we compare the total fusion excitation
functions of the $^{6}$He+$^{59}$Co ({\sc CDCC} calculations) and
$^{4}$He+$^{59}$Co (experimental data of Ref.~\cite{McMahan80}) reactions. The
first calculation (solid line) only includes the reorientation couplings in
fusion without breakup. All continuum and reorientation couplings are included
in fusion with breakup (dashed curve). We observe that both calculated
curves (with and without breakup) give much larger TF cross sections for
$^{6}$He compared to $^{4}$He. We also observe that the inclusion of the
couplings to the breakup channels notably increases the TF cross section for
all energies. 

The same conclusions are reached when we compare on the right panel of Fig.~3
the TF excitation functions of the $^{6}$He+$^{59}$Co ({\sc CDCC} calculations)
and $^{6}$Li+$^{59}$Co (data points from ~\cite{Beck03}, known to be well
described by {\sc CDCC} calculations~\cite{Diaz03}) reactions. For the $^{6}$He
reaction, the incident energy is also normalized with the Coulomb barrier
V$_{B}$ of the bare potential. Extended calculations are in progress to
quantify the role of 1n- and 2n-transfer channels found to be significant in
recent $^{6}$He
data~\cite{Canto06,Raabe04,DeYoung05,Penion06,DiPietro06,Navin06}.

\section{Summary and conclusions}

The {\sc CDCC} method~\cite{Diaz02}, already shown to be rather successful for
fusion~\cite{Diaz03}, can be used to provide the almost complete theoretical
description of all competing processes (total fusion, elastic scattering,
transfer and breakup) in a consistent way. In this paper we have shown that
the $^{6}$Li+$^{59}$Co reaction can be fairly well understood in this
framework although {\sc CDCC} does not separate CF from ICF. The question 
remains open for the halo nucleus $^{6}$He. 

Some of the preliminary {\sc CDCC} results for the $^{6}$He+$^{59}$Co fusion 
process are presented here for the first time. The predictions for the 
$^{59}$Co target are consistent with the data published for other medium-mass
targets such as $^{64}$Zn~\cite{DiPietro06} and $^{63,65}$Cu~\cite{Navin06}.
However a full understanding of the reaction dynamics involving couplings to
the breakup and neutron-transfer channels will need high-intensity radioactive
ion beams to permit measurements at deep sub-barrier energies and precise
measurements of elastic scattering and yields leading to transfer channels
and  to the breakup itself. The application of four-body (required for an 
accurate $\alpha$-n-n description of $^{6}$He) {\sc CDCC} models under current
development~\cite{Matsumoto04,Thompson05} will then be highly desirable.\\ 

\noindent
{\small {\bf Acknowledgments:} One of us (C.B.) would like to thank F. Liang, 
D. Mahboub, A. Moro, N. Rowley, K. Rusek, and V. Zagrebaev for fruitful
discussions.}\\ 

\bibliographystyle{aippprocl}


%


\end{document}